\documentclass[letterpaper]{article} 
\usepackage{aaai25}
\usepackage{times}  
\usepackage{helvet}  
\usepackage{courier}  
\usepackage[hyphens]{url}  
\usepackage{graphicx} 
\usepackage{natbib}  
\usepackage{caption} 
\usepackage{algorithm}
\usepackage{algorithmic}
\usepackage{times}
\usepackage{soul}
\usepackage{url}
\usepackage[utf8]{inputenc}
\usepackage{graphicx}
\usepackage{amsmath}
\usepackage{amsthm}
\usepackage{booktabs}
\usepackage[switch]{lineno}
\usepackage{subcaption}

\usepackage{newfloat}
\usepackage{listings}
\DeclareCaptionStyle{ruled}{labelfont=normalfont,labelsep=colon,strut=off} 
\floatstyle{ruled}
\newfloat{listing}{tb}{lst}{}
\floatname{listing}{Listing}
\title{Importance of User Control in Data-Centric Steering for Healthcare Experts}
\author{
    Aditya Bhattacharya\textsuperscript{\rm 1},
    Simone Stumpf\textsuperscript{\rm 2},
    Katrien Verbert\textsuperscript{\rm 1}
}
\affiliations{
    \textsuperscript{\rm 1}KU Leuven, Belgium\\
    \textsuperscript{\rm 2}University of Glasgow, Scotland, UK\\
    aditya.bhattacharya@kuleuven.be, simone.stumpf@glasgow.ac.uk, katrien.verbert@kuleuven.be
}

\usepackage{bibentry}

\begin{document}

\maketitle

\begin{abstract}
As Artificial Intelligence (AI) becomes increasingly integrated into high-stakes domains like healthcare, effective collaboration between healthcare experts and AI systems is critical. Data-centric steering, which involves fine-tuning prediction models by improving training data quality, plays a key role in this process. However, little research has explored how varying levels of user control affect healthcare experts during data-centric steering. We address this gap by examining manual and automated steering approaches through a between-subjects, mixed-methods user study with 74 healthcare experts. Our findings show that manual steering, which grants direct control over training data, significantly improves model performance while maintaining trust and system understandability. Based on these findings, we propose design implications for a hybrid steering system that combines manual and automated approaches to increase user involvement during human-AI collaboration.
\end{abstract}

%

\section{Introduction}

While Artificial Intelligence (AI) and Machine Learning (ML) have shown significant impact across various applications, high-performing prediction models alone are insufficient for effective human-AI collaboration~\cite{Rezwana2022,cai2019helloai}. Successful collaboration depends on factors beyond model performance, such as user understanding of the system's purpose and its advantages over existing methods~\cite{cai2019helloai}. Limited user control over data instances and the lack of standardised guidelines for incorporating user feedback have further hindered human-AI collaboration~\cite{wondimu2022interactive}. Although prior work has emphasised the need for better integration of end-user feedback~\cite{CHEN2023100780}, the role of user control in these interactions remains a debated topic~\cite{zha2023datacentric,konig2022challenges}. Addressing these gaps is critical for advancing human-centred AI systems, especially for high-stake domains, such as healthcare.

To facilitate human-AI collaboration in healthcare, prior research highlights the benefits of involving healthcare experts in fine-tuning training datasets to improve prediction models~\cite{BhattacharyaCHI2024,bhattacharya2024explanatorydebiasinginvolvingdomain,teso_leveraging_2022}. We refer such an approach as \textit{data-centric steering} as these approaches of finetuning prediction models with user feedback are also aligned with the principles of data-centric AI \cite{Mazumdar2022,zha2023datacentric}. Moreover, healthcare experts’ domain knowledge is particularly valuable for identifying biases and limitations in training datasets, such as those found in patients’ medical records, which could otherwise impact model performance~\cite{bhattacharya2024explanatorydebiasinginvolvingdomain}. Addressing these issues effectively has been shown to significantly improve prediction accuracy and fairness~\cite{bhattacharya2024explanatorydebiasinginvolvingdomain,feuerriegel2020fair}. 

Despite its importance, little research has explored how varying levels of user control during data-centric steering influence model outcomes and collaboration with healthcare experts. This study addresses this gap by presenting two data-centric steering approaches: (1) manual steering and (2) automated steering, which healthcare experts can use for finetuning training datasets. Manual steering grants healthcare experts direct control over training data, enabling them to retain essential data points and predictor variables while removing problematic, corrupt, or irrelevant ones. In contrast, automated steering provides no direct control over the data but uses automated correction algorithms to identify and resolve potential data issues. In automated steering, experts can review these corrections and select methods with a single action, ensuring their consent. 

We conducted a mixed-methods user study with 74 healthcare experts to examine the benefits and challenges of granting greater user control during data-centric steering through an interactive ML system. The process flow of the system is illustrated in Figure \ref{fig:teasure_image}. Through the mixed-methods user study, we aimed to address two key research questions:

\begin{description}
\item [RQ1.] What is the impact of involving healthcare experts in manual and automated steering on model improvement?
\item [RQ2.] How do manual and automated steering influence trust and system understandability for healthcare experts?
\end{description}

By focusing on these questions, we sought to explore the interplay between user control, model performance, and the collaborative dynamics of human-AI systems in healthcare. Considering the findings of this user study, our work has three main contributions relevant to the field of human-centred AI:

\begin{enumerate}

\item \textbf{Empirical Contributions}: Our user study revealed that healthcare experts achieved higher prediction accuracy with manual steering, demonstrating its greater \textit{effectiveness}. Despite the additional effort required during manual steering, it did not adversely affect user trust or system understandability. These findings highlight the potential benefits and trade-offs of empowering healthcare experts with greater control during steering, offering valuable insights into the design of such systems.

\item \textbf{Artifact Contributions}: We designed and developed a data-centric steering system that enables healthcare experts to apply their prior knowledge and configure the training dataset through manual or automated steering methods. The design, source code and technical architecture of our system are open-sourced on GitHub.

\item \textbf{Theoretical Contributions}: Our work presents a set of generic design implications for including healthcare experts in data-centric steering through manual and automated approaches. Additionally, these design implications can be extended to other application domains.

\begin{figure}[h!]
\centering
\includegraphics[width=1.0\linewidth]{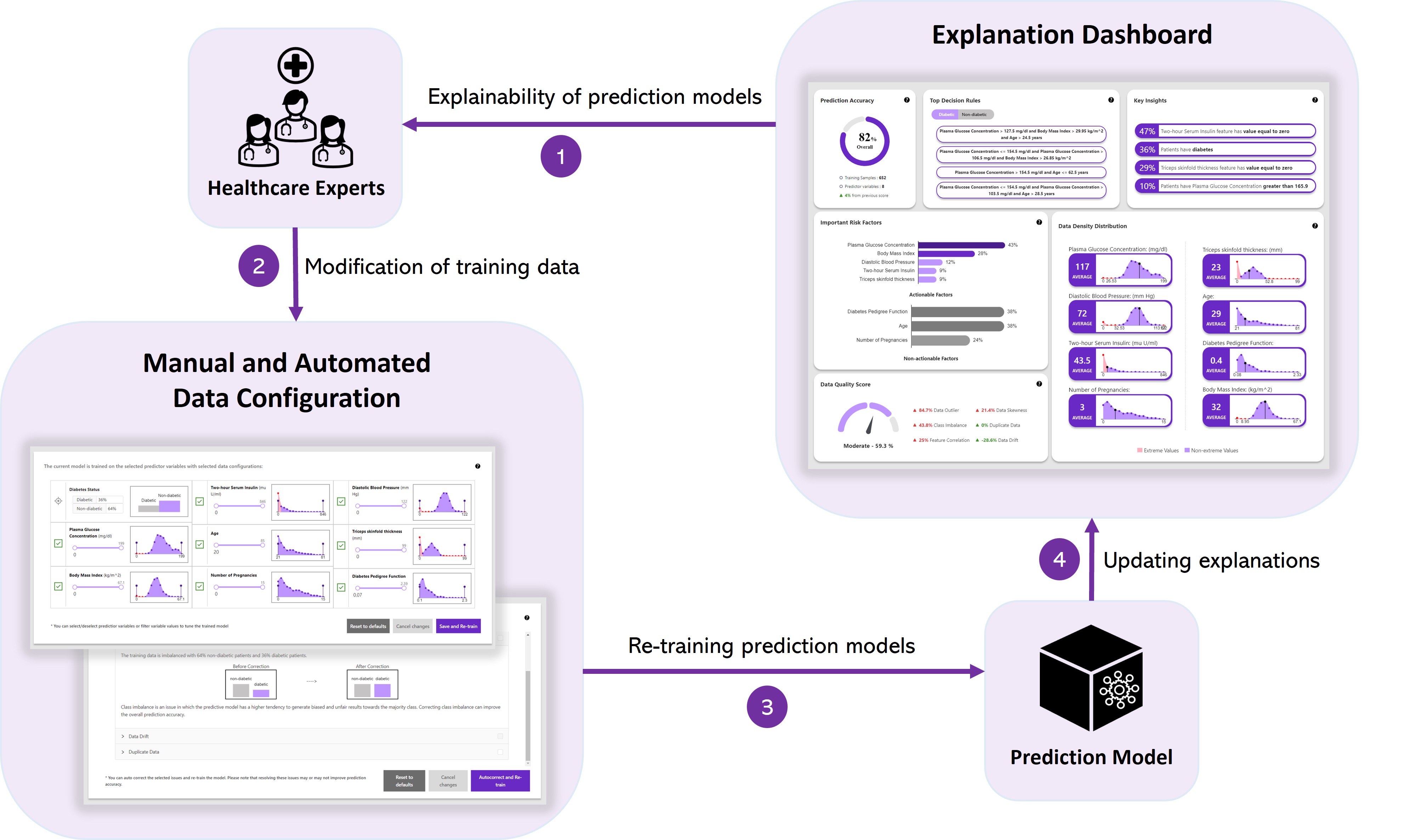}
\caption{ Process flow for our healthcare-focused data-centric steering system. (1) The system consists of a multifaceted explanation dashboard which combines data-centric and model-centric explanations for explaining the working of prediction models to healthcare experts. (2) These explanations can facilitate healthcare experts to share their domain knowledge in modifying the training data for better model performance through manual and automated steering. (3) Prediction models can be re-trained using manual and automated steering on the configured training data. (4) After the prediction model is re-trained, the explanations are regenerated on the updated model and the configured training data.  }
\label{fig:teasure_image}
\end{figure}

\end{enumerate}

\section{Background and Related Work}

\subsection{Data-Centric Steering}

Historically, research in AI/ML has primarily focused on improving models rather than the quality of training datasets~\cite{Mazumdar2022}. While user-in-the-loop methods have been proposed to steer prediction models by adjusting algorithms, parameters and hyperparameters~\cite{teso_leveraging_2022,kulesza_principles_2015,settles2011closing,cakmak2011mixed,kulesza_explanatory_2010}, these \textit{model-centric approaches} face inherent limitations due to issues in the training data~\cite{zha2023datacentric,Mazumdar2022}. Such steering approaches often lead to biased and inaccurate models mainly because of poor data quality~\cite{bhattacharya2024explanatorydebiasinginvolvingdomain,BhattacharyaCHI2024,Mehrabi2021}.
Recently, the growing recognition of the importance of data quality in AI systems has shifted attention toward \textit{Data-Centric AI} (DCAI)~\cite{zha2023datacentric,Mazumdar2022}, driving the development of methods to improve data quality to address issues due to biased, inaccurate and irrelevant predictions.

Feature selection~\cite{Li_2017_feature_selection} is one such method utilised for preparing concise training data by selecting only important predictor variables instead of all the available variables.  Data slicing is another method where a systematic approach is adopted to take a smaller segment of the data instead of the entire data~\cite{Zhang2016}. This method is particularly useful when the predictor variable has high outliers and skewed data distribution. Other methods include quality assessments~\cite{Sadiq_2018} and quality improvements~\cite{Baylor_2017}. Quality assessments and improvement methods involve the detection and correction of common data issues like duplicate records, anomaly data points, correlated features, missing values, data drift detection, biased data, class imbalance and etc~\cite{lones2023avoid,ackerman2022detection,kazerouni2020active,Sadiq_2018,Baylor_2017}. Despite differing sub-goals, these methods share a common objective: improving training data quality for better prediction models~\cite{zha2023datacentric}.

To implement various DCAI methods for model steering, researchers have primarily relied on two approaches: (1) automation and (2) manual user collaboration~\cite{zha2023datacentric}. Automated algorithms are essential for handling the ever-growing volume of data in ML systems. For instance, data generation methods like Random Oversampling~\cite{Menardi2012TrainingAA}, SMOTE~\cite{Chawla_2002}, and ADASYN~\cite{He_2008} address class imbalance by generating synthetic samples for underrepresented categories. Similarly, automated algorithms can detect outliers, handle missing values, remove correlated features and duplicate records \cite{zha2023datacentric}. Automation offer significant advantages, including reduced human error, improved efficiency, higher accuracy, and better reproducibility~\cite{Mazumdar2022}.

However, manual user involvement is critical in tasks where human expertise ensures the alignment of training data with domain-specific expectations~\cite{zha2023datacentric}. For example, manual efforts are vital for tasks like labeling~\cite{zhang2022survey} and filtering data to remove noise or biases~\cite{zha2023datacentric,Zhang2016}. Moreover, these manual and automated approaches are primarily designed for technical AI experts. Acknowledging the importance of involving domain experts with little to no AI knowledge in the AI solution pipeline \cite{bhattacharya2024explanatorydebiasinginvolvingdomain} and to examine the relative strengths of automated and manual approaches in improving training data quality, we conducted experiments comparing both methods with healthcare experts.

\subsection{Explainable AI Methods}
To effectively involve healthcare experts in the process of training, debugging and finetuning ML models, prior work has shown the importance of including Explainable AI (XAI) methods \cite{BhattacharyaCHI2024,Bhattacharya2023,lakkaraju2022rethinking,adadi2018peeking}. Besides increasing the transparency of ``black-box'' algorithms,  explanations can increase the understandability and trustworthiness of ML systems ~\cite{Miller2017,liao2022connecting,Bhattacharya2023}. 

Researchers have categorised explanation methods as \textit{model-centric} or \textit{data-centric}, based on the components of an ML system they address~\cite{BhattacharyaXAI2022}. Model-centric approaches focus on evaluating the importance of parameters and hyperparameters within ML models, such as the SHAP-based feature importance method~\cite{adadi2018peeking,lundberg2017unified}. In contrast, data-centric explanation methods analyse patterns in the training data to justify model predictions~\cite{anik_data-centric_2021}. These methods can summarise data patterns, detect biases and inconsistencies, and explain how issues like data drifts, adversarial attacks, or corrupted features affect model performance~\cite{anik_data-centric_2021,Bhattacharya2023,BhattacharyaXAI2022}.

While both approaches offer unique advantages and limitations, recent studies highlight that combining model-centric and data-centric explanations can provide more comprehensive and effective insights~\cite{BhattacharyaCHI2024,demsar2019}. In our steering approach, we incorporated both types of explanations in a dashboard to support healthcare experts.

\section{Steering System}
\begin{figure}[h!]
\centering
\includegraphics[width=1.0\linewidth]{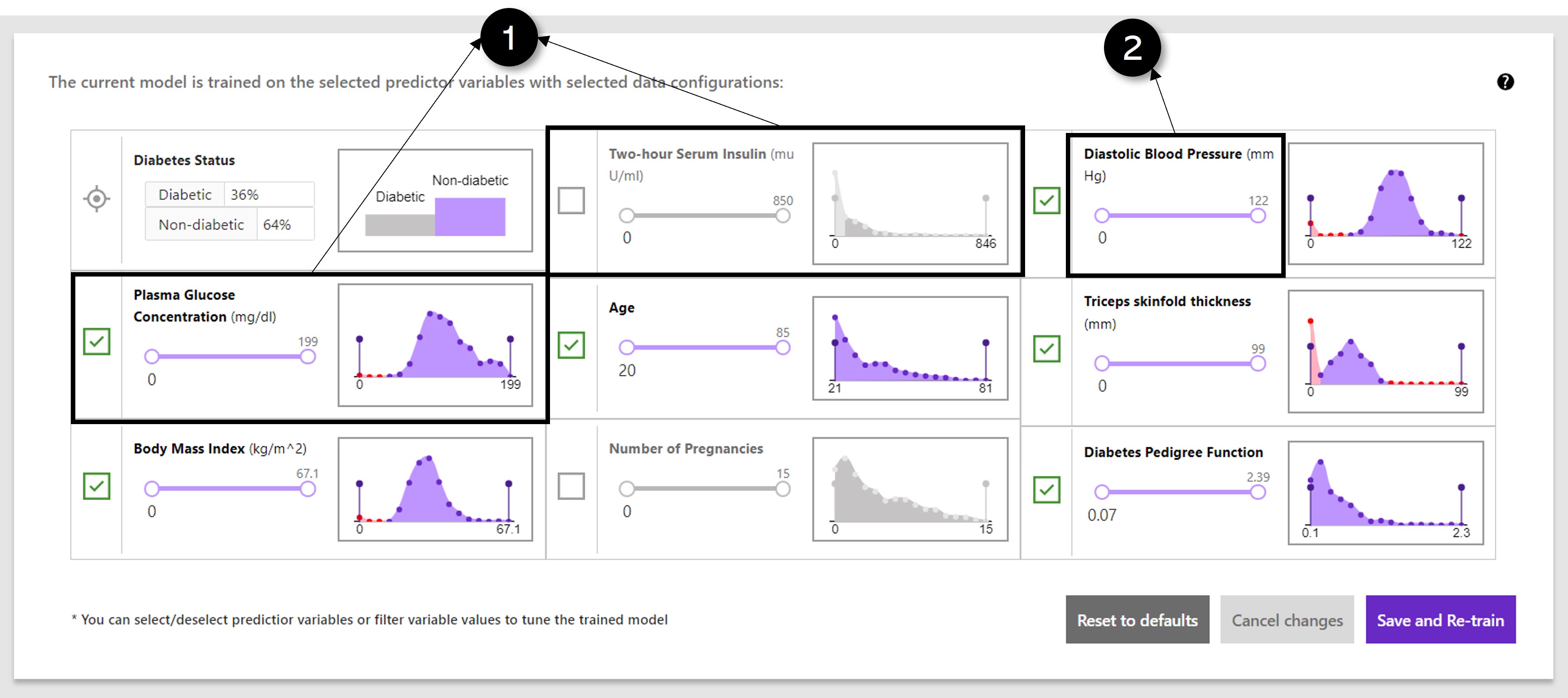}
\caption{ Screenshot of manual steering page that includes (1) feature selection control to include or exclude predictor variables and (2) feature filtering control to set the upper and lower limits for the predictor variables. }
\label{fig:manual_config}
\end{figure}

\begin{figure}[h!]
\centering
\includegraphics[width=1.0\linewidth]{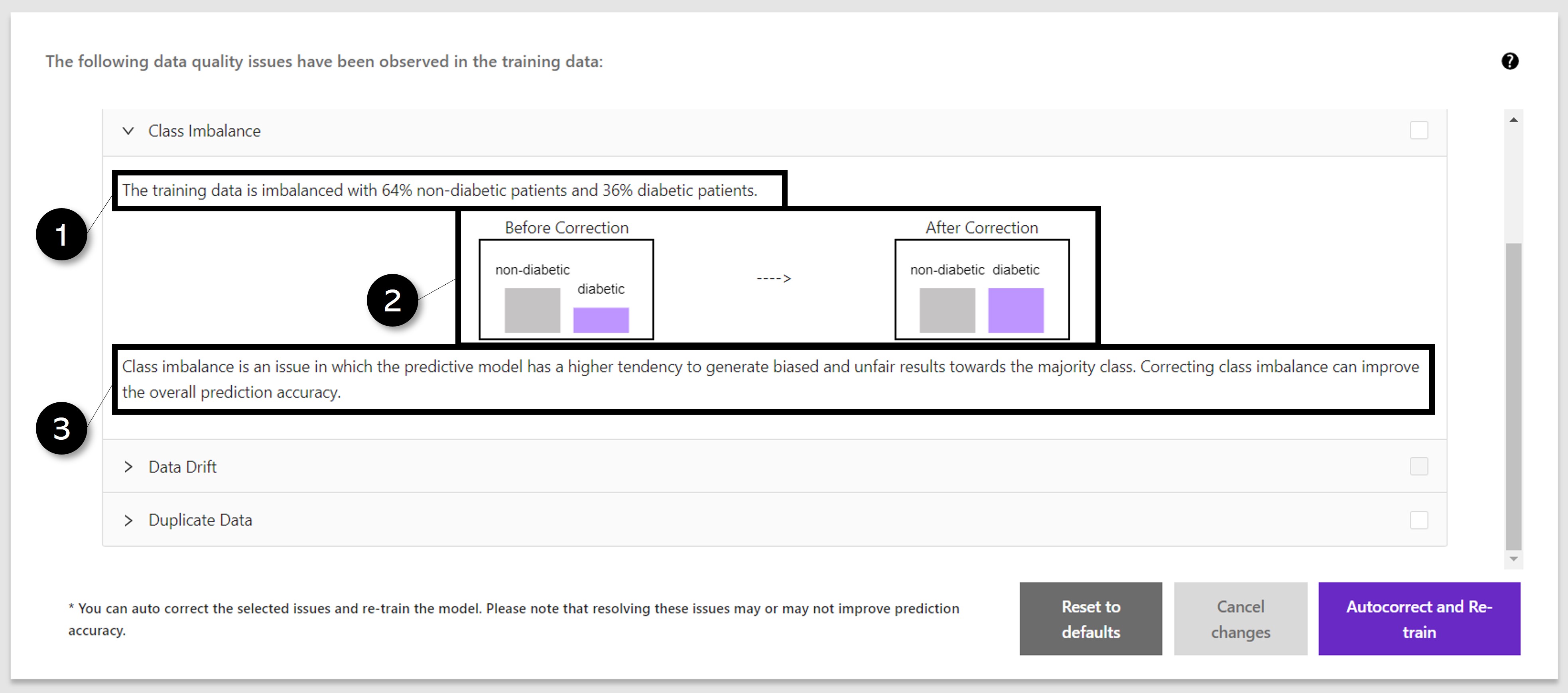}
\caption{Screenshot of automated steering page that includes data issue explanations through (1) the quantified impact of these issues, (2) visualisations displaying before and after correction changes to the data or predictor variables, and (3) description of the issue and how its correction can impact the model performance.}
\label{fig:auto_config}
\end{figure}

\textbf{\emph{System Description}}: Our data-centric steering system includes an explanation dashboard and a configuration page, enabling users to engage in either manual or automated steering (Figures \ref{fig:manual_config} and \ref{fig:auto_config} respectively), but not both simultaneously. It is designed specifically to support healthcare experts, such as doctors and nurses, in steering a diabetes prediction model for early detection of type 2 diabetes.

\begin{figure}
\centering
\includegraphics[width=1.0\linewidth]{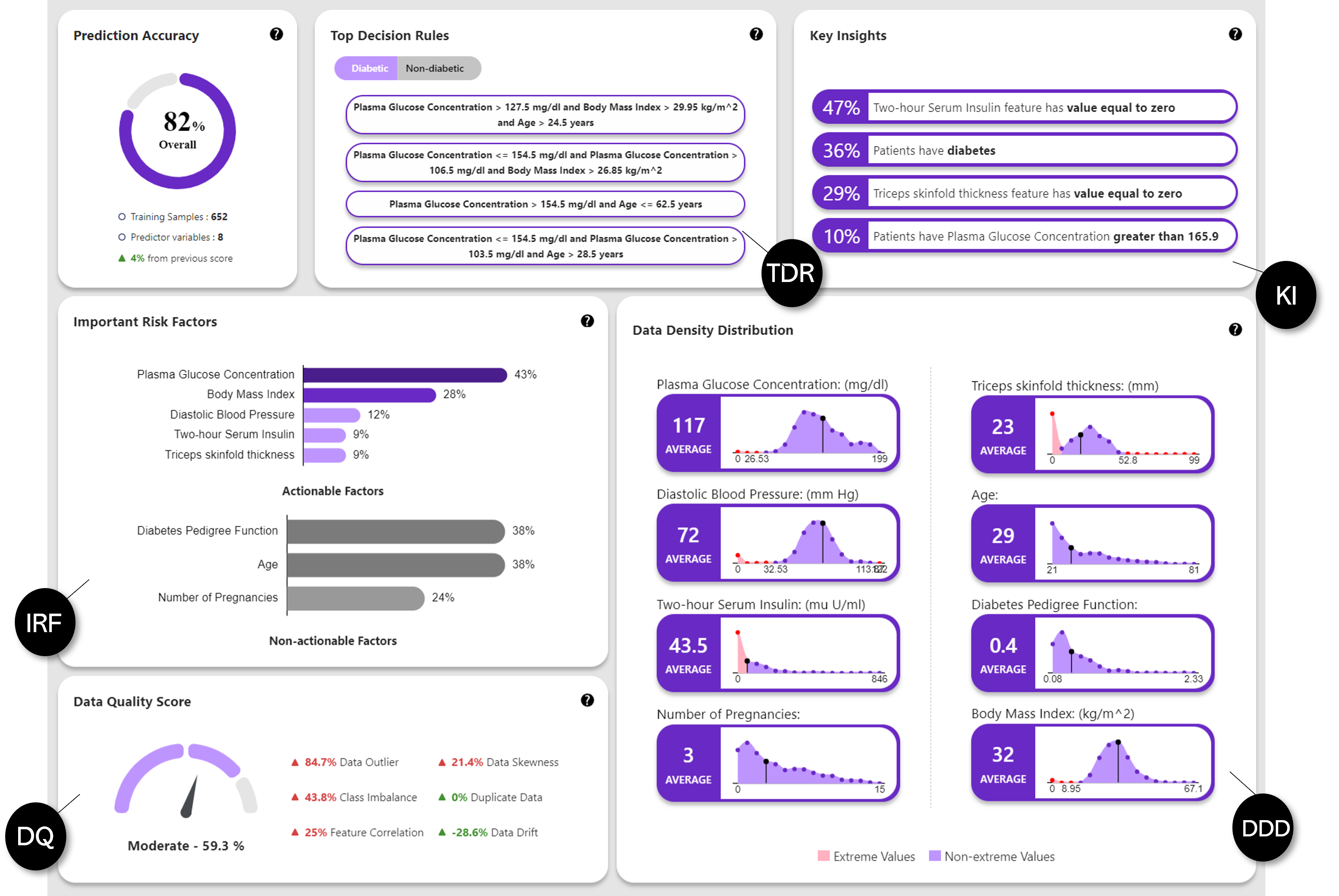}
\caption{Screenshot of the explanation dashboard that includes the following visual components: Top Decision Rules (TDR), Key Insights (KI), Important Risk Factors (IRF), Data Quality (DQ) and Data Density Distribution (DDD).}
\label{fig:explanation_dashboard}
\end{figure}

The design of the explanation dashboard (as shown in fig.\ref{fig:explanation_dashboard}) was inspired by the work of \citeauthor{BhattacharyaCHI2024}, who proposed integrating model-centric and data-centric explanations to improve user guidance in steering models. In line with their framework, the dashboard incorporates several visual components to improve user understanding. These include Top Decision Rules (TDR), which present decision rules generated by surrogate explainers; Key Insights (KI), which provide descriptive statistics from the training data; Important Risk Factors (IRF), which highlight feature importance using SHAP values; Data Quality (DQ), which summarises the estimated quality of the training dataset; and Data Density Distribution (DDD), which illustrates the frequency distributions of predictor variables to offer an overview of the dataset’s characteristics. These elements collectively assist domain experts (like healthcare experts) in understanding and steering prediction systems.

\textbf{\emph{Prediction Model and Dataset}:} The diabetes prediction model was developed using LightGBM algorithm \cite{LightGBM2017}, which was trained a type-2 diabetes detection dataset~\cite{Smith1988-rv}.  This dataset comprises medical records of female patients and includes critical health information such as plasma glucose levels, body mass index, blood pressure, insulin levels, age, number of pregnancies, skinfold thickness, and pedigree function indicating the patient’s family history of diabetes. A detailed data description and the various experiments conducted to train the prediction model are available in the supplementary material.

We selected this dataset for our experiments due to inherent data issues, including class imbalance (with significantly more non-diabetic than diabetic patients), numerous zero-values in predictor variables, and skewed data distributions. These characteristics made it ideal for investigating whether healthcare experts could identify and correct such issues to improve prediction models. Additionally, we avoided overly complex health datasets that could hinder system uderstandability and participant recruitment.

\begin{figure*}
\centering
\includegraphics[width=0.85\textwidth]{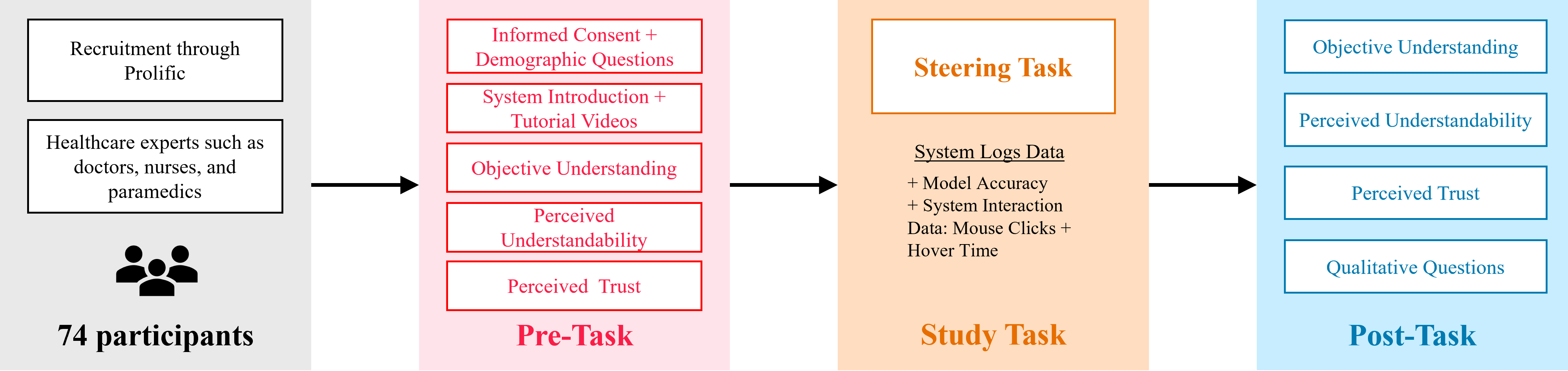}
\caption{User study flow: this diagram illustrates the overall flow of our mixed-methods user study.}
\label{fig:study_flow}
\end{figure*}

\section{Mixed-methods User Study}
\subsection{Study Details}

We conducted a mixed-methods between-subjects user study with 74 healthcare experts to compare manual and automated steering approaches. Participants were recruited from Prolific. On average, participants took 40 minutes to complete the study and were compensated at an hourly rate of \$15 for their time. We have obtained the ethical approval for this study from KU Leuven with the approval number G-2021-4074.

The study included registered and in-training healthcare experts such as doctors, nurses, and paramedics, all of whom had prior experience in treating and caring for diabetic patients. Participants also self-reported familiarity with the predictor variables in the dataset and their relevance for predicting type 2 diabetes. To ensure balanced group assignment, participants were randomly divided into the manual and the automated steering groups, with 37 participants in each. Demographic information for each group is detailed in Table~\ref{tab:participants_info}. 

\begin{table}[h]
\centering
\caption{Participant information for the mixed-methods user study.}
\label{tab:participants_info}
 \scalebox{.68}{
\begin{tabular}{@{}lll@{}}
\toprule
\textbf{}             & \textbf{Manual Group}                              & \textbf{Automated Group}                                    \\ \midrule
NO. OF PARTICIPANTS  & 37                                                        & 37                                                        \\ \midrule
AGE GROUPS           & 20 – 29 years : 29                                              & 20 – 29 years: 31                                              \\
                      & 30 – 39 years: 4                                               & 30 – 39 years: 5                                               \\
                      & 40 – 49 years: 4                                               & 40 – 49 years: 1                                               \\\midrule
GENDER                & Male : 19                                                 & Male : 16                                                 \\
                      & Female : 18                                               & Female : 21                                               \\ \midrule
EDUCATION LEVEL                      & Bachelor: 22                                              & Bachelor: 28                                              \\
                      & Master: 13                                                & Master: 7                                                 \\
                      & Doctorate: 2                                              & Doctorate: 2                                        
                                                           \\\midrule
HEALTHCARE EXPERIENCE & \textless 1 year : 2                                      & \textless 1 year : 2                                      \\
                      & 1 – 3 years : 13                                          & 1 – 3 years : 15                                          \\
                      & 3 – 5 years: 11                                           & 3 – 5 years: 10                                           \\
                      & \textgreater 5 years: 11                                           & \textgreater 10 years: 10 \\
                      
                      
        \bottomrule
\end{tabular}}
\end{table}

\subsection{Procedure}

Participants were provided with detailed instructions outlining the study's objectives, their roles, responsibilities, and rights. After obtaining informed consent and collecting demographic information, we introduced the prototype using tutorial videos. These videos explained the usage scenario and provided an overview of the prediction model, explanation dashboard, and the steering mechanism assigned to each participant. The overall study flow is illustrated in Figure \ref{fig:study_flow}.

Following the tutorials, participants completed pre-task assessments, including objective understanding questions and perceived understandability and trust questionnaires. During this phase, they had full access to the explanation dashboard and their assigned steering method to assist in answering the questions. This step established a baseline for measuring how the steering process influenced their understanding and trust across the manual and automated groups.

Participants then completed a data-centric steering task using their assigned approach. Each participant had 10 minutes to modify the training data and retrain the prediction model, with the ability to configure the data multiple times to maximise model accuracy.

After the steering task, participants completed a post-task questionnaire to evaluate changes in their objective understanding, perceived understandability, and trust in the system. The post-task assessment also included open-ended questions to gather qualitative feedback on their experience. The system evaluation measures recorded during the study are elaborated in the next section.

\subsection{Evaluation Measures}

To address our research questions, we collected the following evaluation measures in our user study. The complete study questionnaires are available in the supplementary material.

\textit{Objective Understanding}: We evaluated participants’ understanding of the system by measuring their objective mental model scores (objective understanding) before and after the steering task. This metric assessed participants’ ability to identify key attributes driving the system’s actions and predict outcomes based on changing conditions, following methods outlined by prior studies \cite{weld2018challenge,kulesza_principles_2015,BhattacharyaCHI2024,Cheng2019}. 

\textit{Perceived Understandability}: In addition to objective understanding, we assessed participants’ self-reported perceived understandability, or their confidence in predicting system behaviour, understanding its decision-making support, and using it effectively without detailed knowledge of its mechanisms \cite{hoffman2019metrics}. This was measured using a questionnaire from  \citeauthor{hoffman2019metrics}, recorded both before and after the steering process.

\textit{Perceived Trust}: Perceived trust, or participants’ confidence in relying on the system, was evaluated using the trust scale from  \citeauthor{Jian2020_trust_scale}, recorded before and after the steering task to observe any changes.

\textit{Post-Steering Model Accuracy}: We measured the updated prediction model accuracy after participants engaged in the steering process, similar to \citeauthor{BhattacharyaCHI2024}, to evaluate whether one group achieved better prediction accuracy improvements.

\textit{Interaction Data}: System interaction data, such as mouse clicks, hover time, and model retraining attempts, were tracked to measure effectiveness and efficiency of manual and automated steering approaches, as per \citeauthor{Verbert2016}. Effectiveness was calculated as the ratio of successful configurations (those that improved accuracy) to total configurations attempted, while efficiency was the ratio of total hover time to successful configurations.

\textit{Qualitative Feedback}: Participants provided qualitative insights into their perceptions of understandability and trust in the system, which were used to identify qualitative factors influencing these perceptions.

\subsection{Data Analysis}

As the quantitative data in our study did not meet normality assumptions, we employed the Mann-Whitney U-test to evaluate statistical significance between groups for the evaluation measures \cite{mccrum-gardner_which_2008}. To assess changes in user understanding and trust before and after the steering task within a particular approach, we used Wilcoxon's signed rank test \cite{mccrum-gardner_which_2008}. Additionally, Spearman's correlation coefficient \cite{mccrum-gardner_which_2008} was calculated to examine relationships between various measures, such as perceived trust, perceived understandability, objective understanding, and post-task prediction accuracy for each group. For qualitative data, we conducted thematic analysis following the approach proposed by \citeauthor{BraunClarkTA} to extract key themes from participant responses. To facilitate comparisons between manual and automated steering approaches, we used a range of plots and graphical visualisations to illustrate differences across the evaluation measures.

\begin{figure*}
\centering
\includegraphics[width=0.99\linewidth]{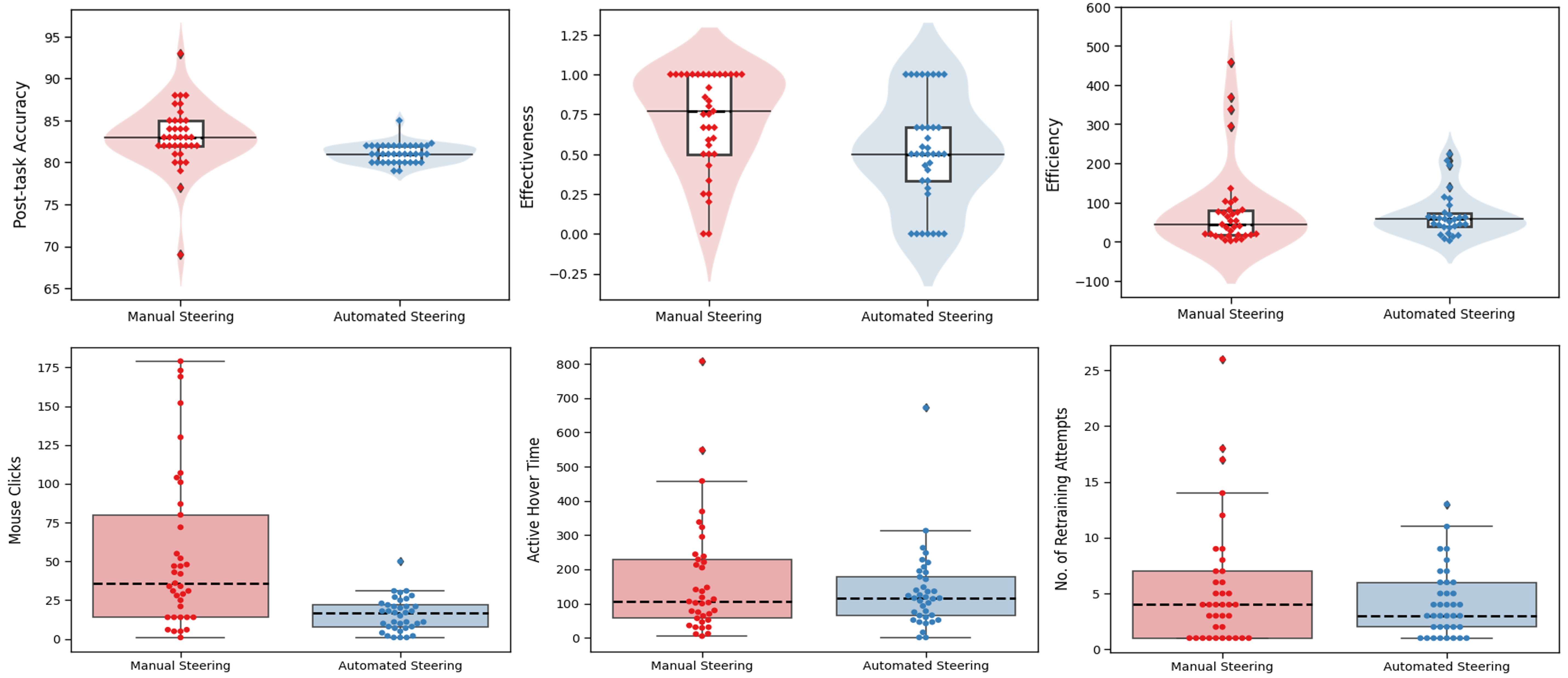}
\caption{(Top row) Boxed-violin plot with marked data points showing the post-task accuracy scores, \textit{effectiveness} and \textit{efficiency} for manual and automated steering groups. (Bottom row) Box-plots showing variations in mouse-clicks, mouse hover time and number of retraining attempts for manual and automated steering groups.
}
\label{fig:post_config_accuracy}
\end{figure*}

\section{Results}

\subsection{What is the impact of involving healthcare experts in manual and automated steering on model improvement? (RQ1)}

The Mann-Whitney U-test revealed that the manual steering group significantly outperformed the automated group in improving prediction accuracy ($U=1054.0, p < .001$). Specifically, 84\% of participants using manual steering achieved higher prediction accuracy than the default, compared to 67\% of those using automated steering. From the system interaction data, manual steering required notably more effort in terms of click counts ($U=1073.0, p < .001$). However, differences in average mouse hover time ($U=706.0, p = .41$) and the number of model retraining attempts ($U=734.0, p = .296$) were not statistically significant.
In terms of \textit{effectiveness}, the manual group achieved a significantly higher score ($0.71$) compared to the automated group ($0.51$, $U=918.0, p = .005$). Conversely, for \textit{efficiency}, the difference between the groups was not statistically significant ($U=480.0, p = .27$). Boxed-violin plots and box-plots in Figure \ref{fig:post_config_accuracy} illustrate the variations between the two groups across these measures, providing insight into RQ1.

\begin{figure}[h]
\centering
\includegraphics[width=1.0\linewidth]{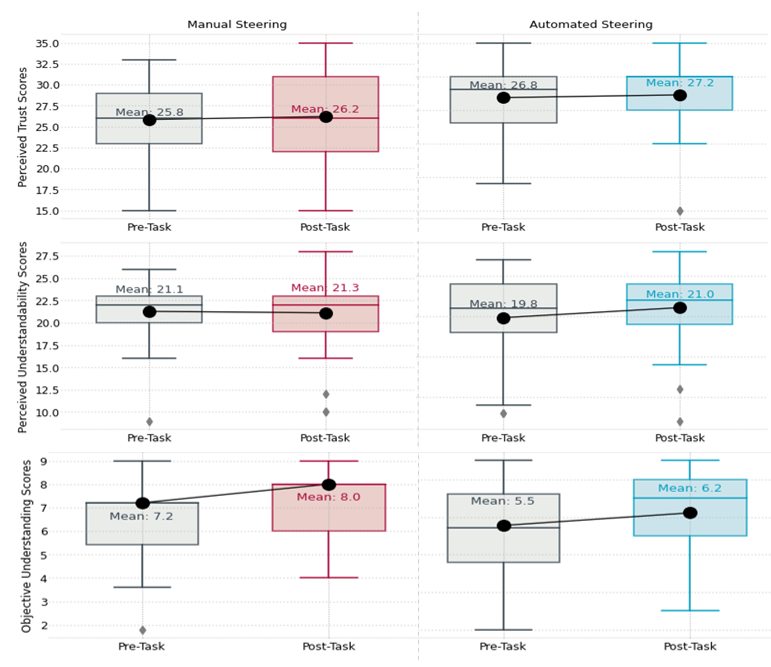}
\caption{Comparing before and after scores of perceived trust,  understandability, objective understanding for manual and automated steering groups.
}
\label{fig:rq2_metrics}
\end{figure}

\subsection{How do manual and automated steering influence trust and system understandability for healthcare experts? (RQ2)}

The change in perceived trust between the manual and automated groups was not statistically significant, as indicated by the Mann-Whitney U-test ($U=672.0, p = .896$). Both groups showed a slight increase in perceived trust after completing the steering task. However, the Wilcoxon signed rank test revealed no significant change in perceived trust from pre-task to post-task for either the manual ($W=220.0, p=.79$) or automated ($W=192.0, p=.40$) groups. Additionally, no significant correlation was found between post-task prediction accuracy and perceived trust for either group (manual: $r=0.21, p=.20$; automated: $r=0.05, p=.68$). Despite the manual group achieving higher model performance, their perceived trust levels were comparable to those of the automated group. Therefore, we conclude that the level of user control in model steering does not significantly affect perceived trust in the AI system.

The change in perceived understandability between the manual and automated groups was not statistically significant, as shown by the Mann-Whitney U-test ($U=263.0, p=.09$). However, the automated steering group showed a greater average increase in perceived understandability (approximately 5\%) compared to the manual group. Despite this, the increase in perceived understandability for the automated group was not statistically significant, as indicated by the Wilcoxon signed rank test ($W=153.5, p=.103$). Similar to perceived trust and understandability, the Mann-Whitney U-test showed no significant difference in the change in objective understanding between the manual and automated groups ($U=646.5, p=.68$). However, the Wilcoxon signed rank test revealed a significant increase in objective understanding for the manual steering group ($W=190.5, p=.019$). In contrast, the increase in objective understanding for the automated group was not statistically significant ($W=263.0, p=.093$). Figure \ref{fig:rq2_metrics} presents box plots comparing the before and after scores of trust and understandability for both groups.

The results indicate that participants in both the manual and automated steering groups had similar levels of trust and understanding. To further explore the impact of control on trust and understandability, we analysed our qualitative data. Thematic analysis revealed key themes that offered deeper insights into how various factors in the steering process influenced user trust and understandability.

\emph{\textbf{Control is important, yet too much control is concerning}}: Both groups valued the ability to configure the training data, which enhanced their trust in the model’s results. While most manual steering participants gained trust through hands-on adjustments, some expressed concerns about maintaining data integrity. One participant noted: ``\textit{There is too much risk involved when dealing with medical data directly. Incorrect changes can be too risky.}'' However, manual steering also helped participants better understand how changes to variables influenced predictions: ``\textit{After playing around with the configurations, I have a better understanding of the predictions.}''

\emph{\textbf{Explanations about data quality and data issues increase trust and reliability}}: Both groups emphasised the importance of high-quality data for accurate predictions. Trust in the system grew when the explanation dashboard showed improved data quality after the steering process. As one participant stated: ``\textit{The data quality score gives me more confidence to trust the predictions. If it’s lower, I’ll be more sceptical and rely on my own judgment.}'' Several also suggested that understanding the data collection process would further enhance trust: ``\textit{If the data quality is high, the system is more reliable, but knowing about the data source would increase my trust.'}'

\section{Discussions}

\subsection{Limitations}\label{subsec_limitations}
Before discussing the broader implications of our work, we acknowledge the following known limitations to ensure the transparency of our research:
\begin{enumerate}     
    \item \textit{Potential limitations in the experimental prototype:} There are certain areas for potential improvement in our experimental prototype. Although the dataset used to train the prediction model fits all our experimental requirements, future research should consider investigating the impacts of different steering approaches on larger datasets. Additionally, integrating the explanation dashboard and the data configuration screens into the same view could enhance the system's usability and user interaction experience. 
    \item \textit{Unexplored impact of manual and automated steering for other use cases}: The implications of this research work are limited only to classification models trained on structured datasets. Other use cases involving different types of datasets, such as image or text data, might require a different approach for manual and automated steering and were outside the scope of this research.
\end{enumerate}

\subsection{Can We Depend Only on Manual Steering?}

Our findings support providing healthcare experts with greater control through manual steering to achieve more \textit{effective} model improvement. However, the manual approach comes with limitations. It increases the risk of human errors, particularly when users have only a partial understanding of the system. Such user-induced errors can degrade training data quality, which in turn affects model accuracy and diminishes user trust. Thus, our research raises an essential open question: ``\textit{What takes precedence for healthcare experts during data-centric steering: the pursuit of higher prediction accuracy or the assurance of better data quality?}'' We argue that systematic collaboration between domain experts (like healthcare experts) and AI specialists is vital to balancing accuracy with data quality during manual steering. Such collaboration leverages domain knowledge alongside technical expertise, fostering informed decision-making that improves both model performance and the integrity of the data. We echo the recommendations from prior researchers \cite{BhattacharyaDebiasing2025, ZhangCHI20202} who emphasised the importance of establishing a complementary partnership between domain experts and AI experts, especially in high-stakes domains, for building fair and responsible AI systems.

\subsection{Towards Hybrid Steering: Combining Manual and Automated Approaches}

Manual and automated steering methods are not mutually exclusive in real-world applications. Given the distinct advantages observed in both approaches, we propose a hybrid system that integrates manual and automated steering to accommodate domain experts (like healthcare experts) who prefer varying levels of control. While our experimental setup did not incorporate hybrid steering as the primary objective of our study was to explore the comparative differences between manual and automated steering, future implementations should consider combining both approaches to maximise the benefits for healthcare experts.



\subsection{Involving a Group of Healthcare Experts During Manual Steering in a Practical Setting}

Certain participants from the manual steering group expressed scepticism when their attempts to improve the model led to degraded performance instead of improvement. Such significant drops in prediction accuracy could adversely affect their trust in the system. To address this problem, we suggest replacing individual data-centric steering with a peer-approval and group consensus process similar to \citeauthor{BhattacharyaCHI2024} Leveraging the collective knowledge of a team of healthcare experts has the potential to overcome the limitations of individual feedback, such as human bias and errors in training data \cite{Mehrabi2021}. By combining their expertise, a group of healthcare experts can achieve a more comprehensive understanding of the system and greater confidence in performing manual steering effectively. Moreover, involving groups of healthcare experts in steering sessions can help distribute cognitive load, making the process more manageable in time-constrained settings. In practice, we envision these sessions taking place collaboratively with AI experts, who can provide technical guidance and help mitigate any unintended effects on the prediction model resulting from domain expert interventions.

\subsection{Improving Inter-Stakeholder Collaboration in ML Systems Using Data-Centric Steering}

ML systems often involve a variety of stakeholders with different backgrounds, such as developers, business leaders, policymakers, and users, who typically operate in isolation~\cite{preece2018stakeholders}. While prior research has highlighted the importance of fostering collaboration across these groups~\cite{preece2018stakeholders,Nahar2022,teso_leveraging_2022,bhatt2020machine}, the dynamics of collaboration between ML experts and domain specialists remain underexplored. Our findings suggest that both manual and automated steering approaches can play a key role in improving inter-stakeholder collaboration within ML systems. Building on \citeauthor{BhattacharyaCHI2024}'s guidelines, we propose a peer-approval process for steering systems that involves diverse stakeholders. This process leverages their collective technical expertise, domain knowledge, and practical experience to improve training data quality and develop more robust and contextually relevant ML models.

\subsection{Design Implications for Data-Centric Steering Systems For Domain Experts}
To summarise the main findings and observations from our study with healthcare experts, we share the following implications for tailoring steering systems for domain experts:
\begin{itemize}
    \item \textit{Hybrid Steering for Balanced Control and Automation}: A hybrid approach that integrates manual and automated steering can help domain experts balance control and efficiency. While manual steering allows for domain expertise-driven adjustments, automated assistance can minimise human errors and improve workflow efficiency.
    
    \item \textit{Facilitating Peer-Approval and Group Consensus}: Instead of relying solely on individual expert feedback, incorporating peer-review mechanisms among domain experts with diverse knowledge and background can improve the reliability of data-centric steering. This collaborative approach mitigates biases and errors, leading to more robust model adjustments.

    \item \textit{Enhancing Explainability Through Interactive Visualisations}: Domain experts require clear explanations of model changes and their impact on predictions. Step-by-step interactive visualisations and explanations of training data quality that highlights the underlying issues can improve understanding and trust in the steering process.

    \item \textit{Inter-Stakeholder Collaboration for Improved Data Governance}: AI systems for high-stakes domain (like healthcare) involve multiple stakeholders, including ML engineers, policymakers, and domain experts. Facilitating structured collaboration through shared decision-making frameworks can improve training data quality and model relevance.

    \item \textit{Implementing Rollback and Version Control Mechanisms}: Inspired by version control systems like GitHub, steering systems should include rollback features that allow healthcare experts to track, revert, and audit changes. This ensures transparency and accountability in manual data modifications for domain experts.

\end{itemize}

Even though this work primarily focuses on understanding the perspective of healthcare experts, these implications are transferable to other application domains that require extensive domain knowledge for the development of responsible AI systems. We encourage future researchers working in human-centred AI to empirically evaluate these preliminary implications and refine them into concrete frameworks for robust data-centric steering systems for all domain experts with limited AI knowledge.

\subsection{Beyond Model Performance Improvement}

While our study primarily focused on how data-centric steering impacts model performance, we believe its benefits extend well beyond accuracy metrics. Involving healthcare experts in the steering process can also support tasks such as bias detection and mitigation, as well as aligning training data more closely with real-world clinical scenarios. Prior work on AI bias and fairness \cite{BhattacharyaDebiasing2025, Mehrabi2021, jamainternmed, feuerriegel2020fair} highlights the importance of user involvement and domain expertise in addressing these broader concerns. Additionally, expert-guided data-centric steering can aid in the debugging and auditing of AI systems, helping ensure their compliance with ethical and societal standards established for responsible AI deployment \cite{2019DataQA, liao2022humancenteredexplainableaixai, masis2023interpretable, szymanski2025human}. We envision that combining data-centric steering with explainable AI techniques can not only reduce model errors and bias, but also contribute to the development of more transparent, trustworthy, and fair AI systems.

\subsection{Future Work}
Future research could investigate the combined effects of manual and automated steering. Additionally, conversational AI, such as chatbot-based interactions, could also improve system explainability and assist healthcare experts in creating more effective data configurations. To deepen understanding, future studies should engage diverse stakeholder communities to examine how manual and automated steering approaches influence inter-stakeholder collaboration. Prior research underscores the importance of rollback mechanisms, \citeauthor{kulesza_principles_2015} advocating for the reversible principle and  \citeauthor{BhattacharyaCHI2024} highlighting the need to track user changes in interactive ML systems. Similar to version control systems like GitHub, incorporating such features is strongly recommended to improve the usability and practicality of data-centric steering systems for domain experts.

\section{Conclusion}
In conclusion, our research examined the potential for healthcare experts to improve prediction models using both manual and automated data-centric steering approaches. We developed a steering system for healthcare experts to guide a diabetes prediction model and conducted a mixed-methods user study with 74 participants. The study provided a detailed analysis of the benefits of each steering approach across multiple evaluation metrics. Our findings show that manual steering led to significantly higher prediction accuracy, demonstrating greater effectiveness. While manual steering required more effort, it did not substantially affect participants' trust or understanding of the system. Based on these results, we recommend granting more control to healthcare experts in fine-tuning prediction models. However, for practical applications, a hybrid steering approach that combines the strengths of both manual and automated methods would be ideal.

\section{Concluding Reflections}

\subsection{Positionality Statement}

As researchers situated at the intersection of HCI and AI, we recognise that our perspectives are shaped by our technical backgrounds and institutional affiliations with access to advanced computational resources. While we involved healthcare professionals to ground our work in real-world practices, we are not clinicians ourselves. This outsider perspective influences our interpretations of needs and system usability for clinical practices. We acknowledge the importance of continued interdisciplinary collaboration to more accurately reflect the experiences of our target users.

\subsection{Ethical Considerations Statement}

This study was conducted in accordance with the ethical guidelines of our institution and was approved through a multistage review process by the institutional ethics committee. The study design was carefully developed to uphold ethical standards from the outset. All participants provided informed consent prior to participation, and no identifiable data was collected; all responses were anonymised to ensure privacy. We took measures to minimise participant burden and emphasised their right to withdraw from the study at any time. In developing the steering system, we prioritised transparency and explainability to support responsible AI use in healthcare. Future work will continue to uphold these ethical principles, with a focus on inclusive and accountable system design.

\subsection{Adverse Impact Statement}

While our work aims to improve human-AI collaboration in healthcare through data-centric steering, we acknowledge potential risks associated with these approaches. Granting domain experts direct control over training data may inadvertently introduce biases or lead to overfitting, particularly in the absence of rigorous validation by AI experts. Moreover, enabling user-driven steering could increase the system's vulnerability to adversarial manipulation or unintended misuse. To mitigate these risks, we recommend that such steering systems be deployed within controlled environments, with oversight from both AI and domain experts. Broad deployment should be preceded by thorough testing on representative datasets and ongoing evaluation to prevent the amplification of existing inequities or other unintended harms.

\bibliography{aaai25}

\end{document}